\documentclass{PoS}
\usepackage{amsmath}
\usepackage{txfonts}
\usepackage{longtable}
\title{Multi-Frequency VLBI Observations of the Gravitational Lens B2016+112}

\ShortTitle{Multi-Frequency VLBI Observations of the Gravitational Lens B2016+112}

\author{\speaker{Anupreeta More}\\
Max-Planck-Institut fuer Radioastronomie\\
        E-mail: \email{anupreeta@mpifr-bonn.mpg.de}}

\author{Richard Porcas\\
Max-Planck-Institut fuer Radioastronomie\\
        E-mail: \email{porcas@mpifr-bonn.mpg.de}}

\abstract{We present Global VLBI and HSA images of the gravitational lens
B2016+112 at 18, 6 and 3.6~cm. Previous VLBI observations showed that 
images A and B (which are clearly lensed images of a single background
source) and the elongated region C are each divided into components.  Our
new high-resolution maps reveal more components in images A and B,
clearly demonstrating their expected opposite parities.  According to the
scenario of Koopmans et al. (2002), the arc-like region C consists
of two merging, partial images (``C1-C2'') of just a small region of the
same background source, seen with high lens magnification.  We have
determined the spectra and relative positions of the components
within all four images in order to test this scenario.
We find that the outer north-west components in images A and B
do indeed have radio spectra similar to the components seen
in C1 and C2.
}

\FullConference{The 8th European VLBI Network Symposium on New Developments in VLBI Science and Technology
                and EVN Users Meeting  \\
                September 26-29 2006\\
                Torun, Poland}

\begin{document}
\section{Introduction}
The gravitational lens system B2016+112, 
discovered in 1981 in the MIT-Green Bank 6-cm survey,
shows three radio components A, B and C (see Figure~1 left).
The overall
integrated spectrum of this system is that of a Gigahertz Peaked
Spectrum source.
Components A and B, separated by 3.4 arcseconds, are associated with
the lensed images of an AGN at redshift $z=3.273$ (\cite{law84} \cite{sch85}
\cite{sch86}; see Figure~1, centre). The radio and optical positions of A and B are coincident within the
measurement errors and they also have similar spectra in both radio and optical, as
expected for lensed images.
Koopmans et al (\cite{koo02})
describe the optical spectrum as that of a type-II quasar.

Component C, which is the brightest radio source in the field,
lies 2 arcseconds away to the south-east of B
and is extended in the E-W direction.
It is associated with a diffuse extended object in the infrared.
A giant elliptical galaxy D at $z=1.01$ (\cite{sch85}), lying close
to the centroid of A, B and C, observed in \textit{K} and \textit{I}
bands (\cite{law93}), acts as the primary lens (see Figure~1, right).
Because the radio spectrum of C is flatter than those of A and B, and there
is no associated optical nucleus, it is not clear whether this is also
lensed emission from the same background source.

\begin{figure}[ht]
 \begin{minipage}[ht]{0.3\textwidth}
   \begin{center}
   \includegraphics[width=4.7cm]{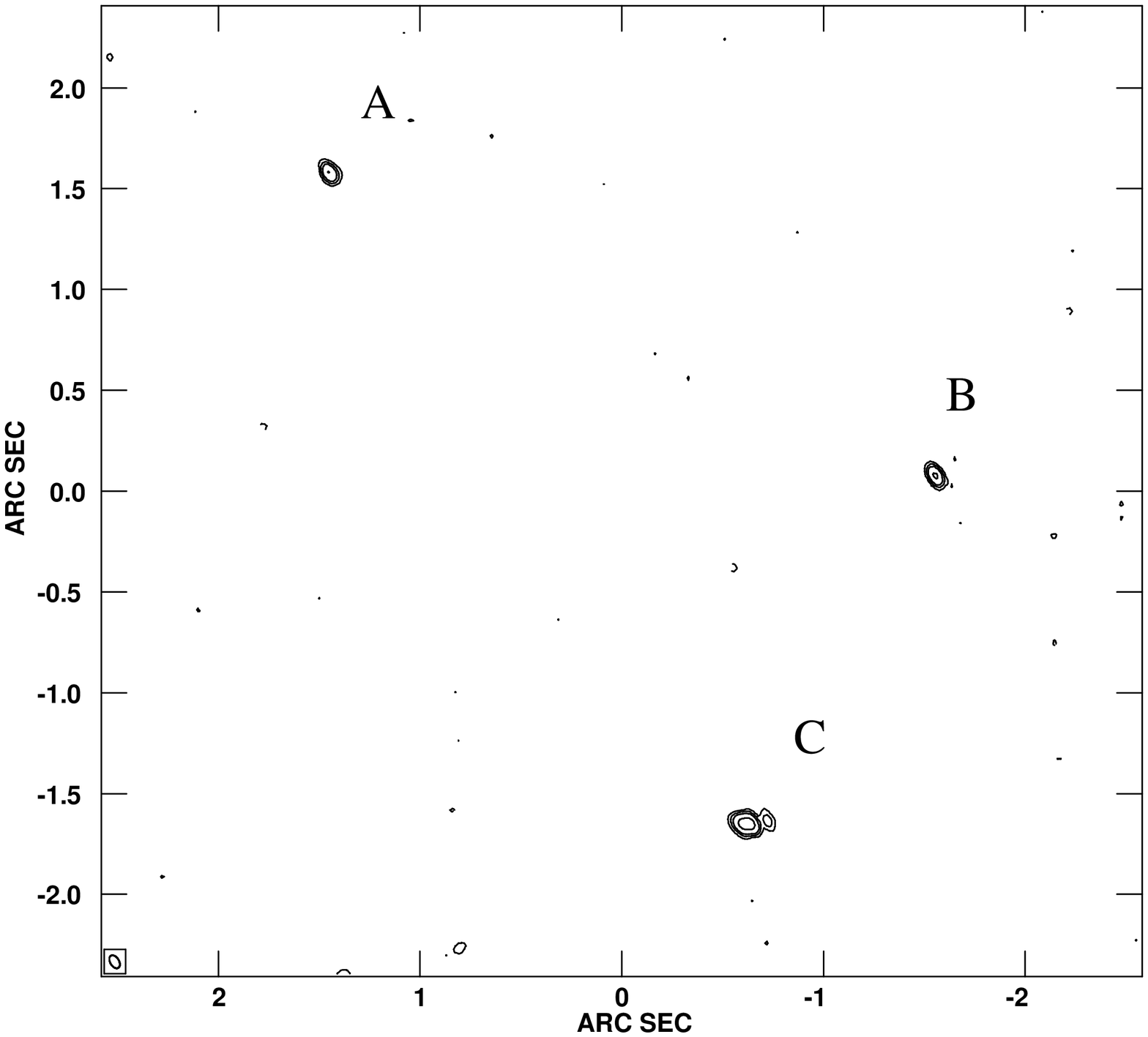}
   \end{center}
  \end{minipage}
\hfill
 \begin{minipage}[ht]{0.3\textwidth}
   \begin{center}
   \includegraphics[width=4.7cm]{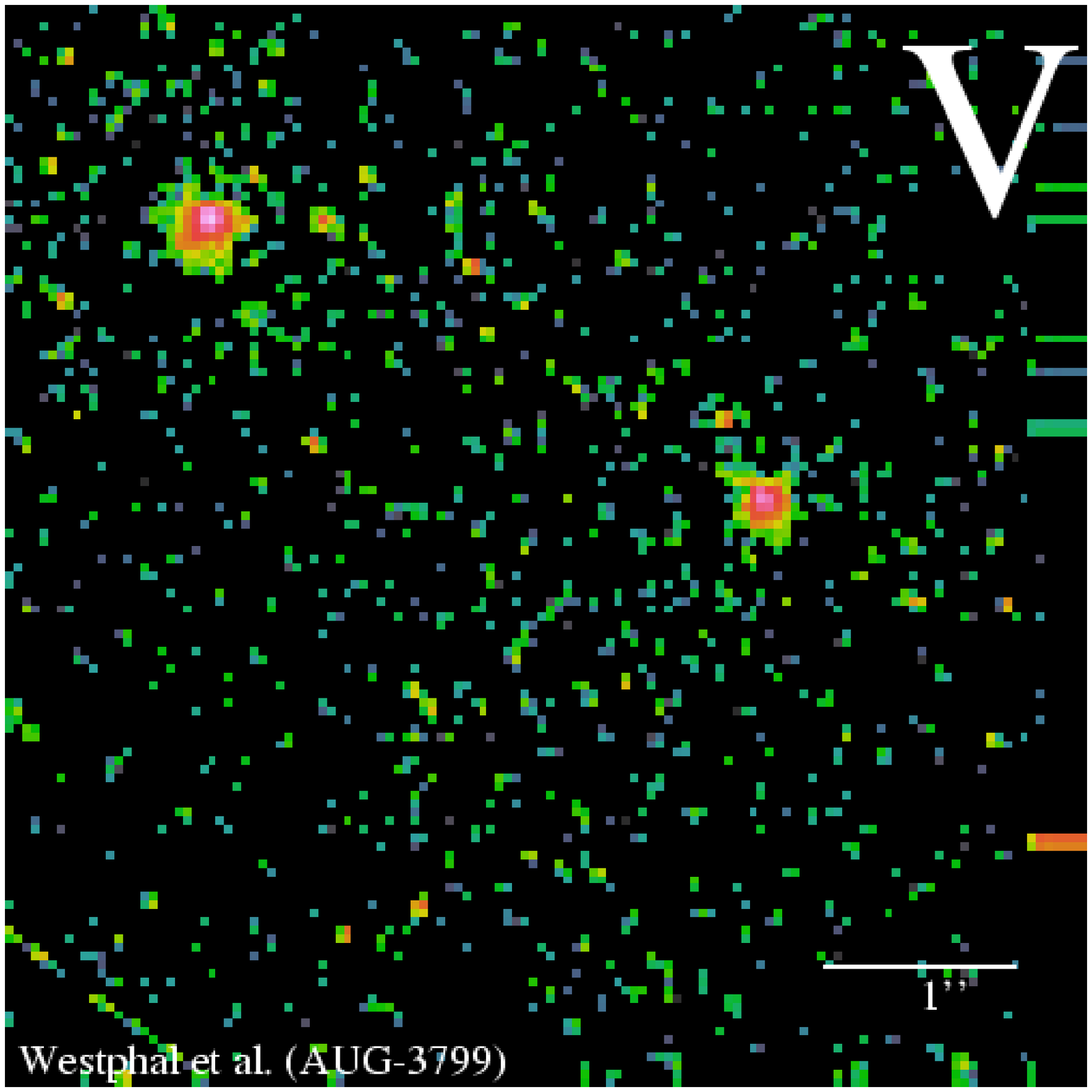}
   \end{center}
  \end{minipage}
\hfill
 \begin{minipage}[ht]{0.3\textwidth}
  \begin{center}
   \includegraphics[width=4.7cm]{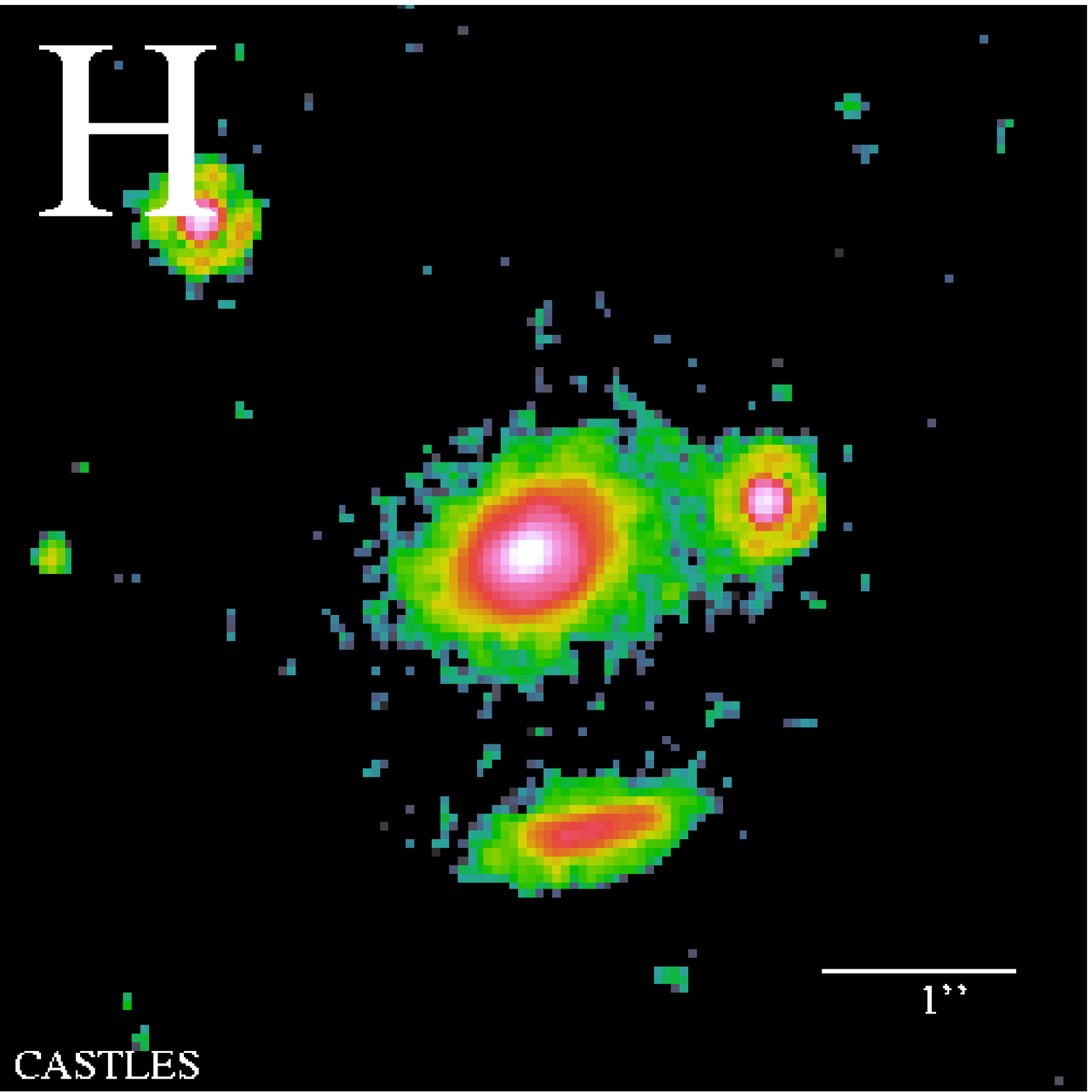}
  \end{center}
 \end{minipage}
\caption{\textit{Left}: MERLIN map of B2016+112 at 6~cm showing A, B and C;
 C is resolved into two sub-components.
\textit{Centre:} HST V-band image showing A and B lensed images of the optical nucleus.
\textit{Right:} HST H-band image showing the lensing galaxy D (z~=~1.01) in addition
to A, B and C. HST images taken from the CASTLES website.
}
\end{figure}

Images A and B are each further resolved into two components as shown
in 18~cm-VLBI observations (\cite{hef91}) and C is resolved into four
components from 6~cm-MERLIN and 18~cm-EVN observations (\cite{gar94},
\cite{gar96}). These show that the outer two components in region C
have steeper spectra than the inner two components.

The lens model of this system by Koopmans et al. (\cite{koo02})
consists of a single screen which contains the primary lens D, two
singular isothermal sphere (SIS) mass distributions and an external
shear.
In this model B2016+112 is interpreted as a quadruply imaged
lens system, in which the diamond caustic crosses the radio source
such that just part of the ``counter-jet'', host galaxy and narrow
line regions are imaged at A, B and C (where C consists of two partial
images, C1 and C2).  The complete source, including all radio
components and the optical nucleus, is only doubly imaged (A and B).
This model predicts unusually high magnification ($\sim$ 300) in
region C.  Support for this interpretation is provided by spectral
observations showing that C has features at the same redshift as A and
B.  In this work, we analyze the spectra of all the radio
components in order to test Koopmans' scenario.

\section{Observations and Data Reduction}
Multi-frequency Global VLBI and HSA observations were made of B2016+112
(see Table~1). B2029+121 was used for phase calibration and B2134+004 as a fringe-finder.
The data were processed at the VLBA
correlator, producing 16 frequency samples per baseband channel.
\begin{table}[!ht]
\begin{center}
\label{Table 1}
\begin{tabular}{c c c c c c c}
\hline Date & Wavelength  & Antennas  & Obs. time & Integ.  &  BB chans. &  Polarization\\
   &     &           &  (hours)  & time(sec) & per polsn. &     \\ \hline
 25Feb02 & 18~cm   & Eb, Jb, Mc, On, Tr & 17    &   2    &   4x8MHz   &  LCP-only      \\
  &         & VLBA, Y, Ro, Go        &   &  &  &\\
  17Nov01 &   6~cm   & Eb, Jb, Mc, On, Wb &   17    &   1    &  2x8MHz   &  LCP+RCP  \\
  &         & VLBA, Ar                &   &  &  & \\
  30Apr06 & 3.6~cm   & VLBA, Eb, Gb, Ar     &   7    &   1    &    4x8MHz   &  LCP+RCP  \\\hline
\end{tabular}
\caption{Observational details}
\end{center}
\end{table}

Further processing of the datasets was carried out with the NRAO
\textsc{aips} package.  No averaging of the data in either time or
frequency was made, in order to reduce the effects of bandwidth and
fringerate smearing when imaging this wide field lens system.  First,
the a priori amplitude calibration derived from system temperature
measurements and antenna gains was applied to the data.  Then the
instrumental phase, delay and rate solutions determined from
fringe-fitting the B2029+121 data were applied to make a map of
B2016+112.  Three sub-fields centered on A, B and C were imaged.  This
phase referenced map was then used as an initial model for phase
self-calibration of the lens data. Several runs of the tasks
\textsc{calib} and \textsc{imagr} were performed to refine the phase
and amplitude solutions before a satisfactory map was obtained.  All
maps were made using a weighting scheme between uniform and natural
(parameter \textsc{robust}~=~0 in the task \textsc{imagr}) to give a
reasonable balance between high resolution and high sensitivity.  In
order to determine the flux densities of components in the maps,
single Gaussian models were fitted using task \textsc{jmfit}; where
this was not possible the integrated flux was estimated using
\textsc{imstat} or \textsc{tvstat}.

\section{Results and Discussion}
The maps at all 3 wavelengths are shown in Figure 2 (images A and B)
and Figure 3 (region C).  In addition to the 2 components noted by
Koopmans et al (A1, A2, B1, B2) we identify in these higher resolution
maps 3 new components (labelled 3, 4 and 5) in both A and B.  The
non-collinear line of the 5 components clearly demonstrates the
opposite image parities of A and B.

The 4 main components previously noted in region C are
clearly seen at all wavelengths.
The labelling of components in region C has been changed here to reflect
the obvious symmetry
between the left and right sides,
and its interpretation as 2 merging images. Partial image C1 has
components C1-1 and C1-2; their counterparts
in C2 become C2-1 and C2-2.

Figure~4 $\it{(left)}$ shows the spectra of all 5 components in A (the
corresponding components in B are essentially the same) and
$\it{(right)}$ the 2 components in both partial images C1 and C2.  All
exhibit steep spectra except A2 (and B2), C1-2 and C2-2, which show
flattened spectra between 18 and 6~cm. Following a standard
description of extragalactic radio sources we would identify A2
\begin{figure}
   \begin{center}
   \includegraphics[width=6.5cm]{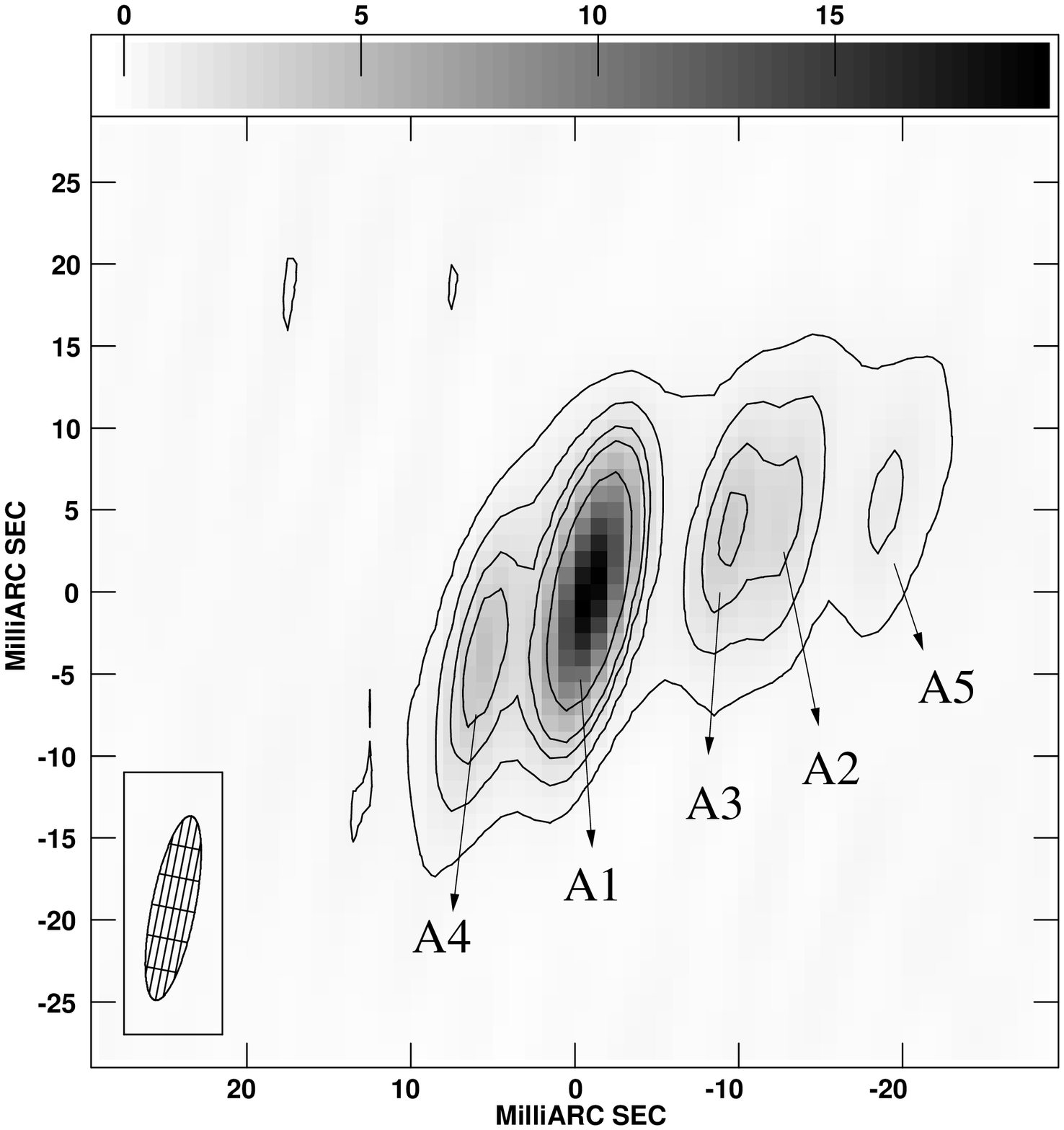}
   \includegraphics[width=6.5cm]{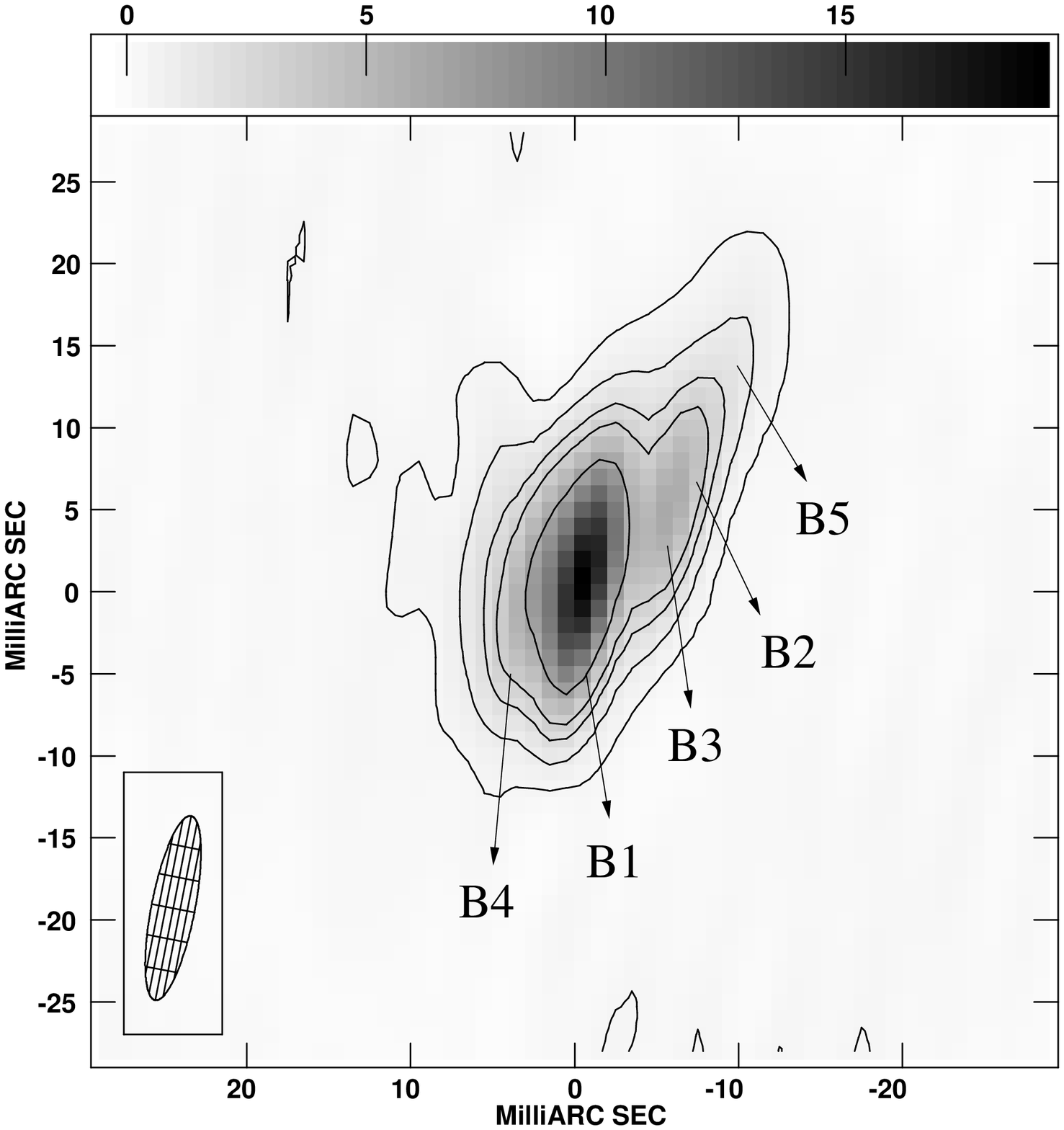}
   \includegraphics[width=6.5cm]{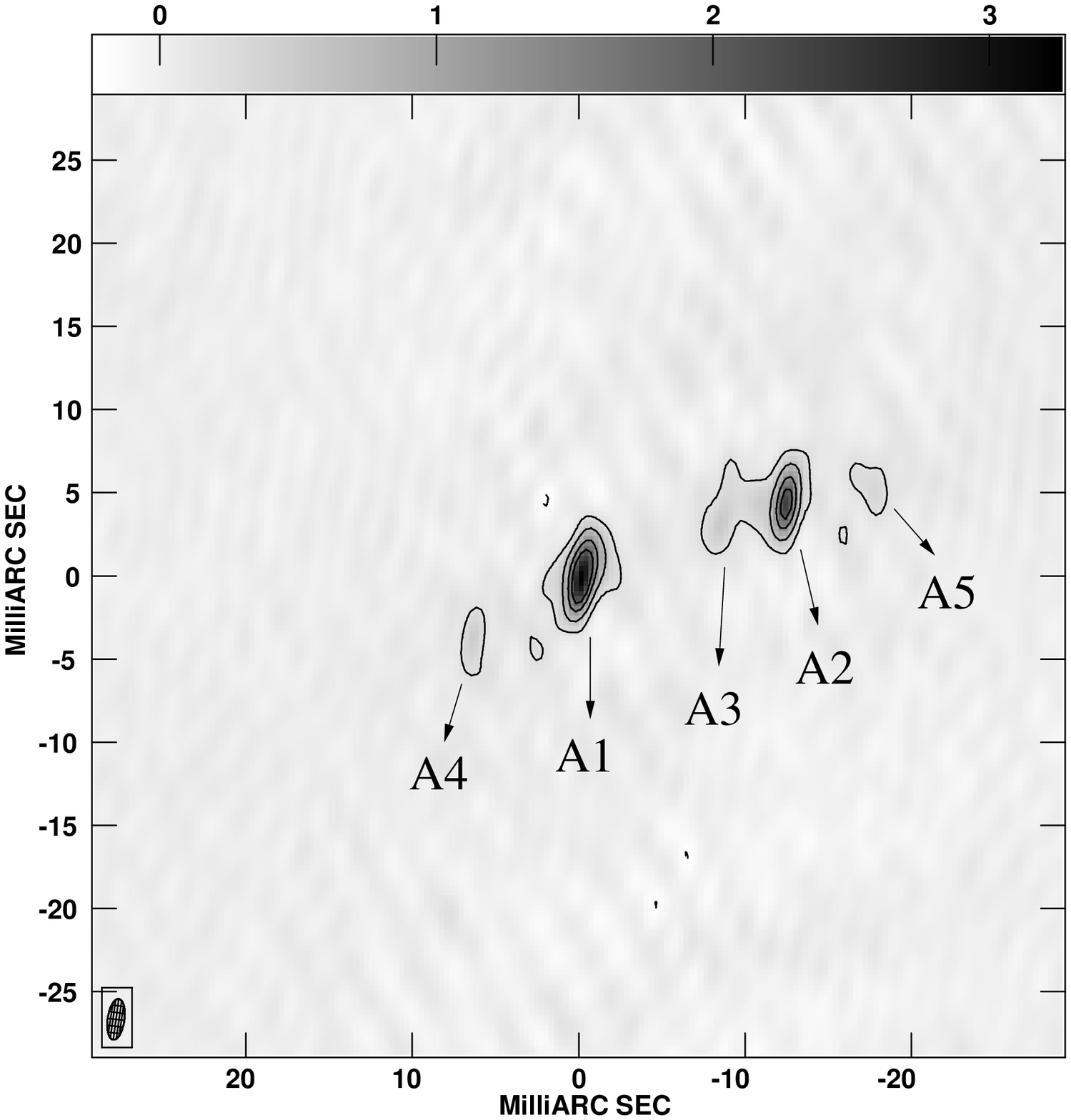}
   \includegraphics[width=6.5cm]{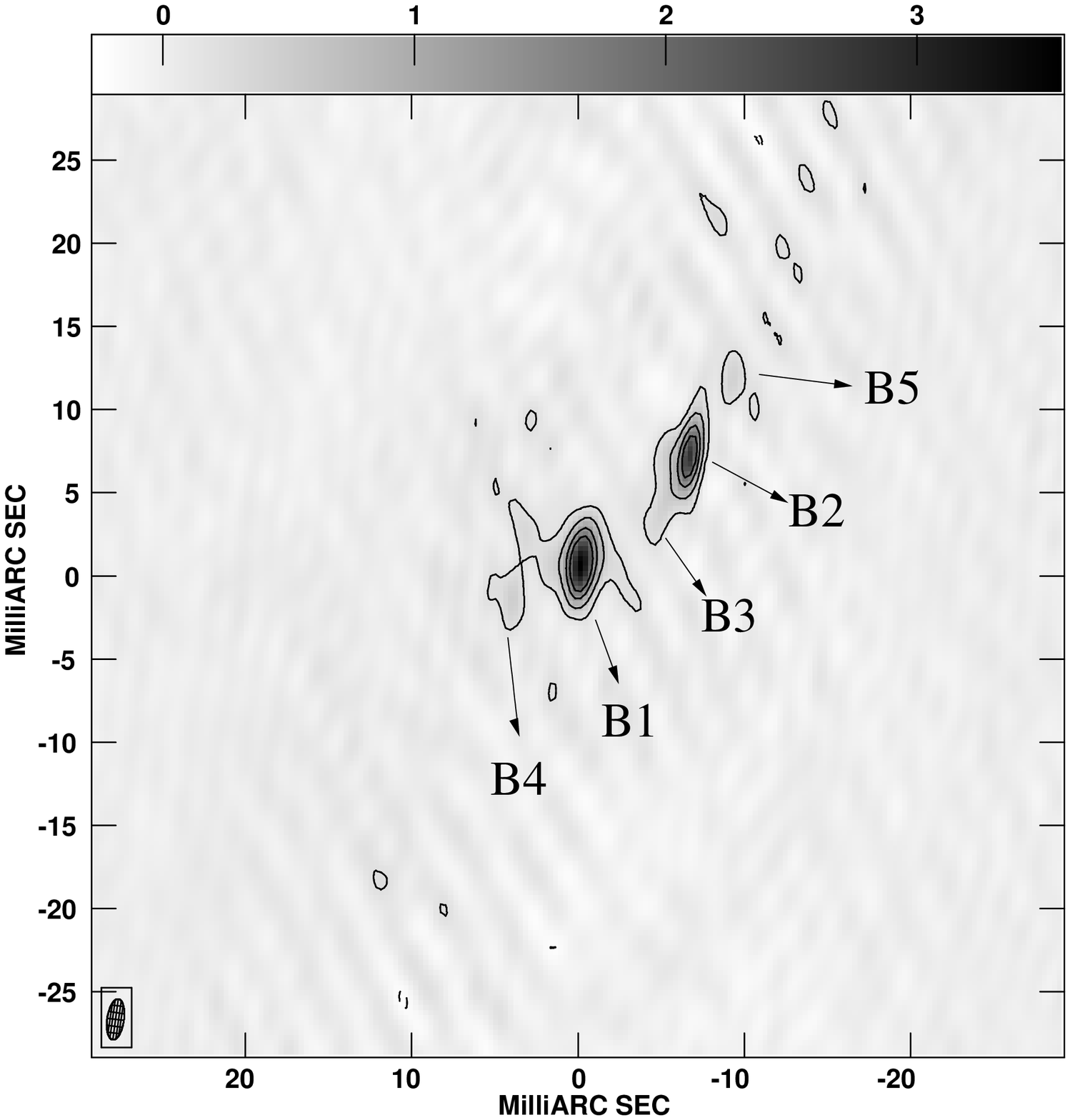}
   \includegraphics[width=6.5cm]{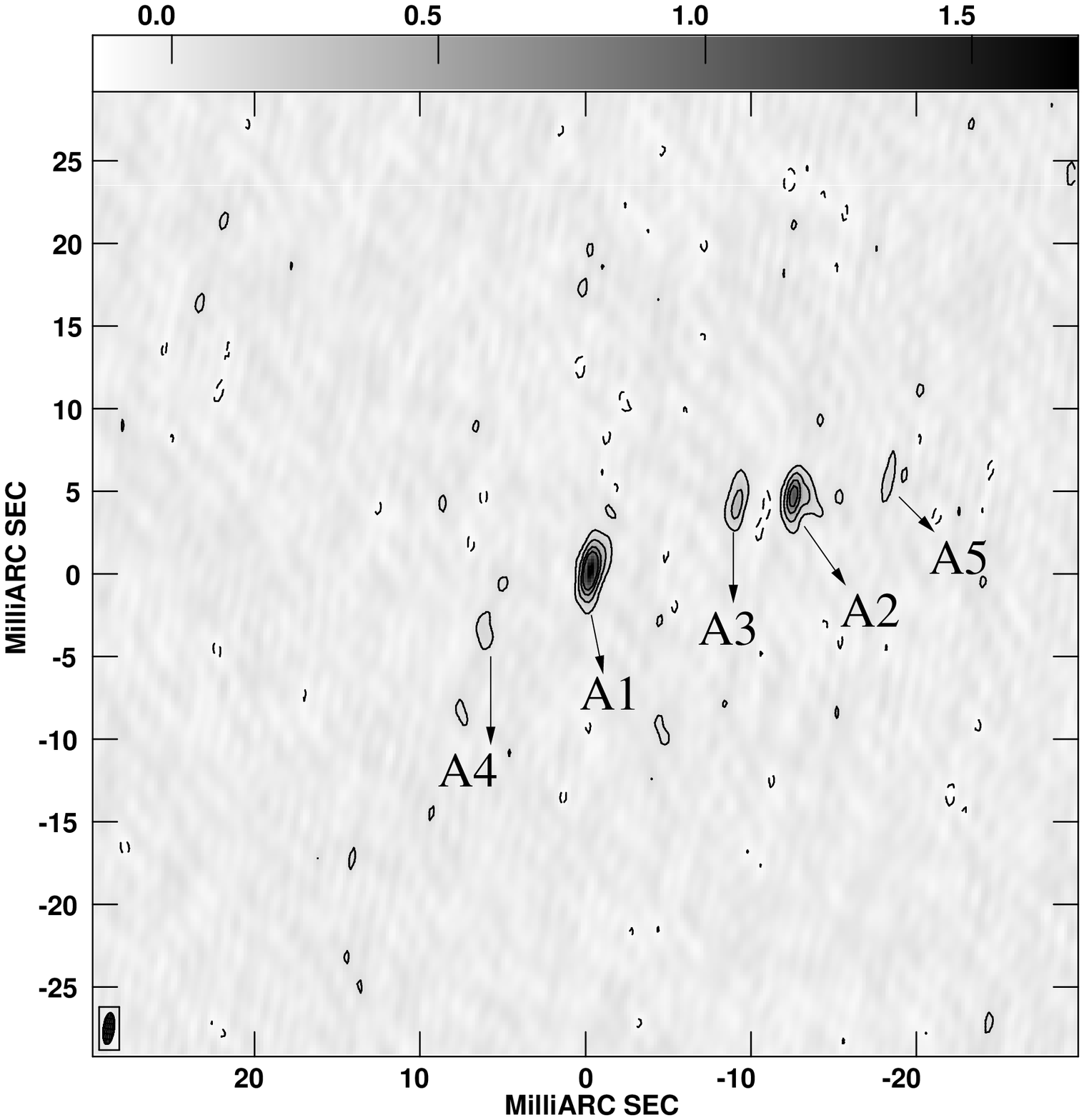}
   \includegraphics[width=6.5cm]{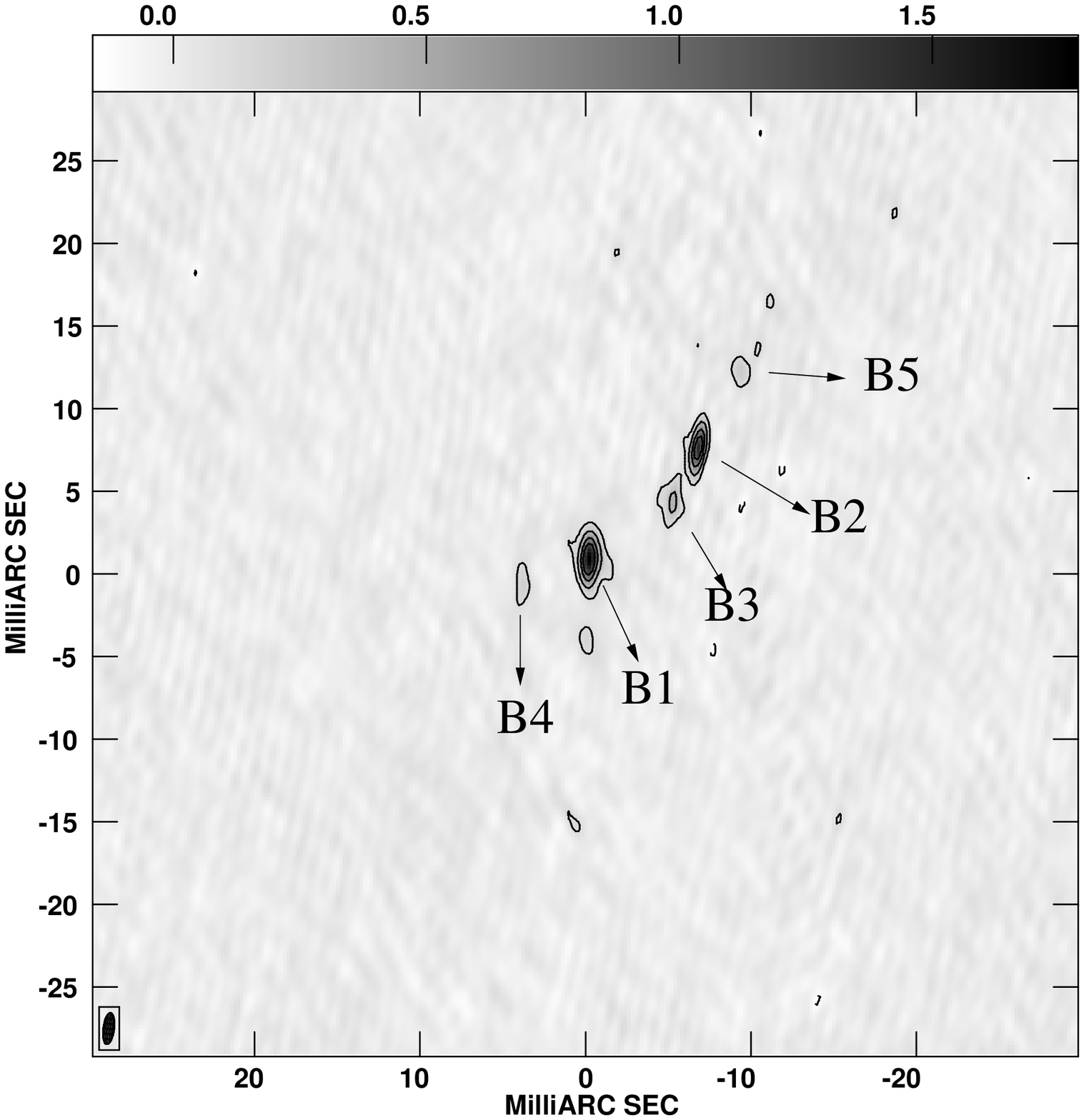}
   \end{center}
\caption{Each row shows $\it{(left)}$ image A and $\it{(right)}$ image B
at (from top to bottom)
18~cm ($\sigma$~=~0.12~mJy~beam$^{-1}$),
6~cm ($\sigma$~=~0.07~mJy~beam$^{-1}$) and
3.6~cm ($\sigma$~=~0.04~mJy~beam$^{-1}$).
The contours
are (-3, 3, 9, 18, 27, 54) $\times$ $\sigma$ in the respective maps.
The restoring beams used were
(18~cm) $11.1\times2.6$~mas$^{2}$ and PA~=~-~10.58$^{\circ}$,
(6cm) $2.4\times1$~mas$^{2}$ and PA~=~-~8.03$^{\circ}$ and
(3.6cm) $1.8\times0.6$~mas$^{2}$ and PA~=~-~7.02$^{\circ}$ .
}
\end{figure}
\begin{figure}
 \begin{minipage}[t]{1\textwidth}
   \begin{center}
   \includegraphics[height=5cm,width=12cm]{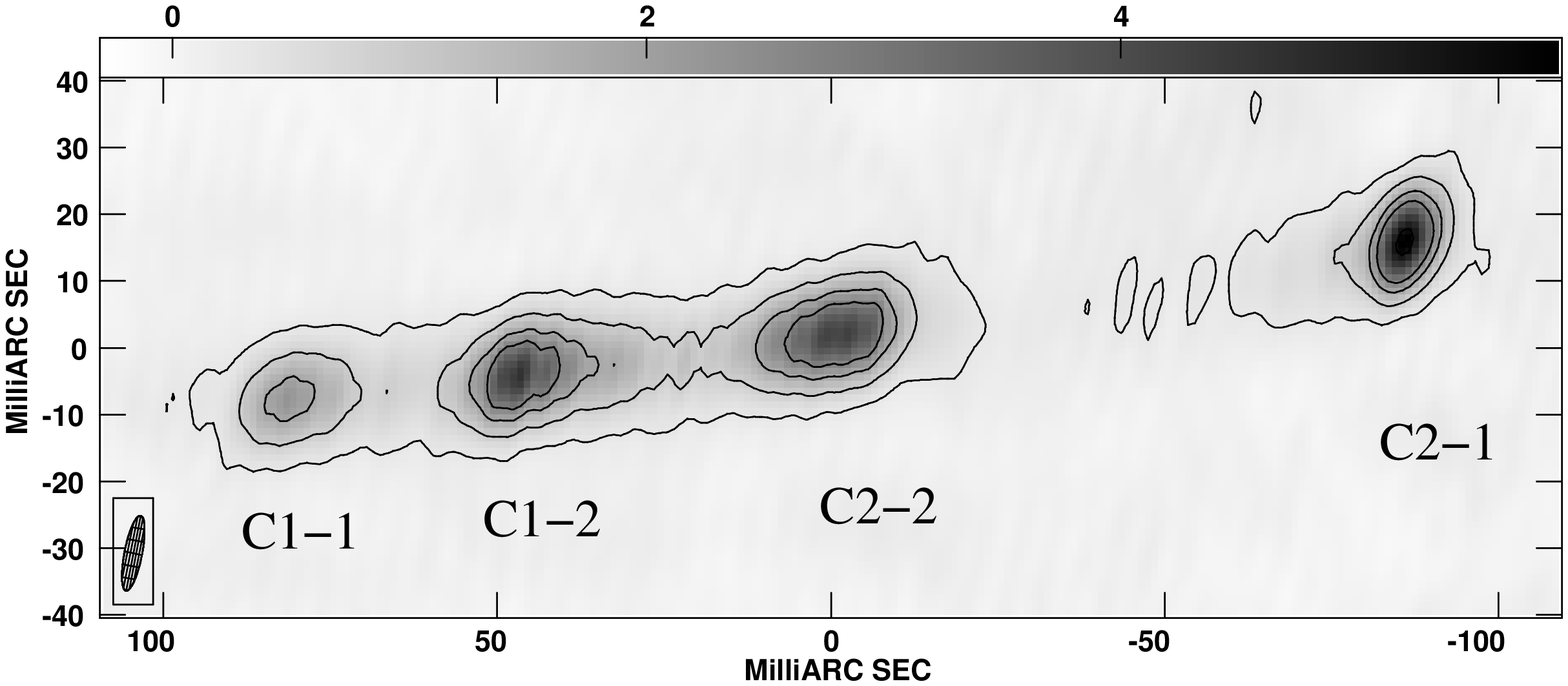}
   \end{center}    
  \end{minipage}
\end{figure}
\vspace{0.5cm}
\begin{figure}
 \begin{minipage}[t]{1\textwidth}
  \begin{center}
   \includegraphics[height=5cm,width=12cm]{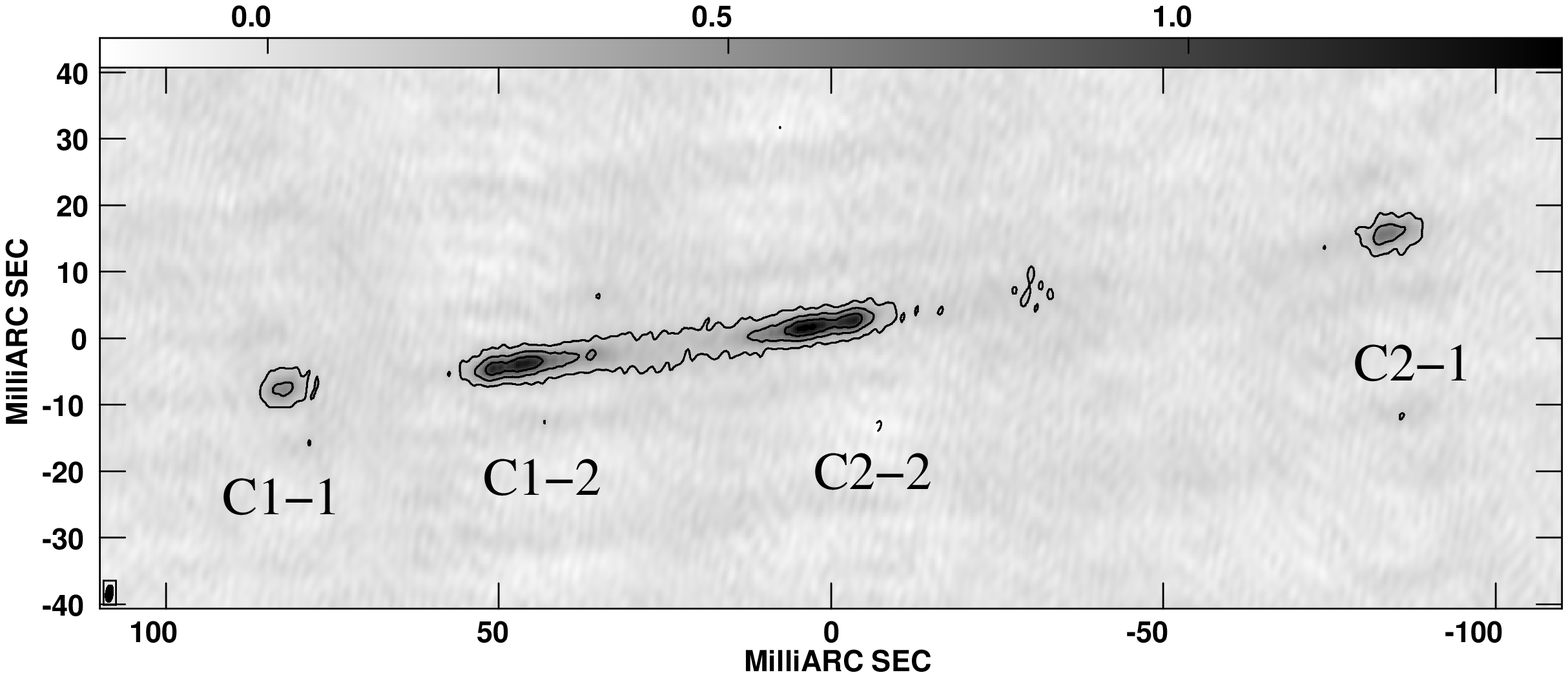}
   \end{center}
 \end{minipage}
\end{figure}
\vspace{0.5cm}
\begin{figure}
 \begin{minipage}[t]{1\textwidth}
  \begin{center}
   \includegraphics[height=5cm,width=12cm]{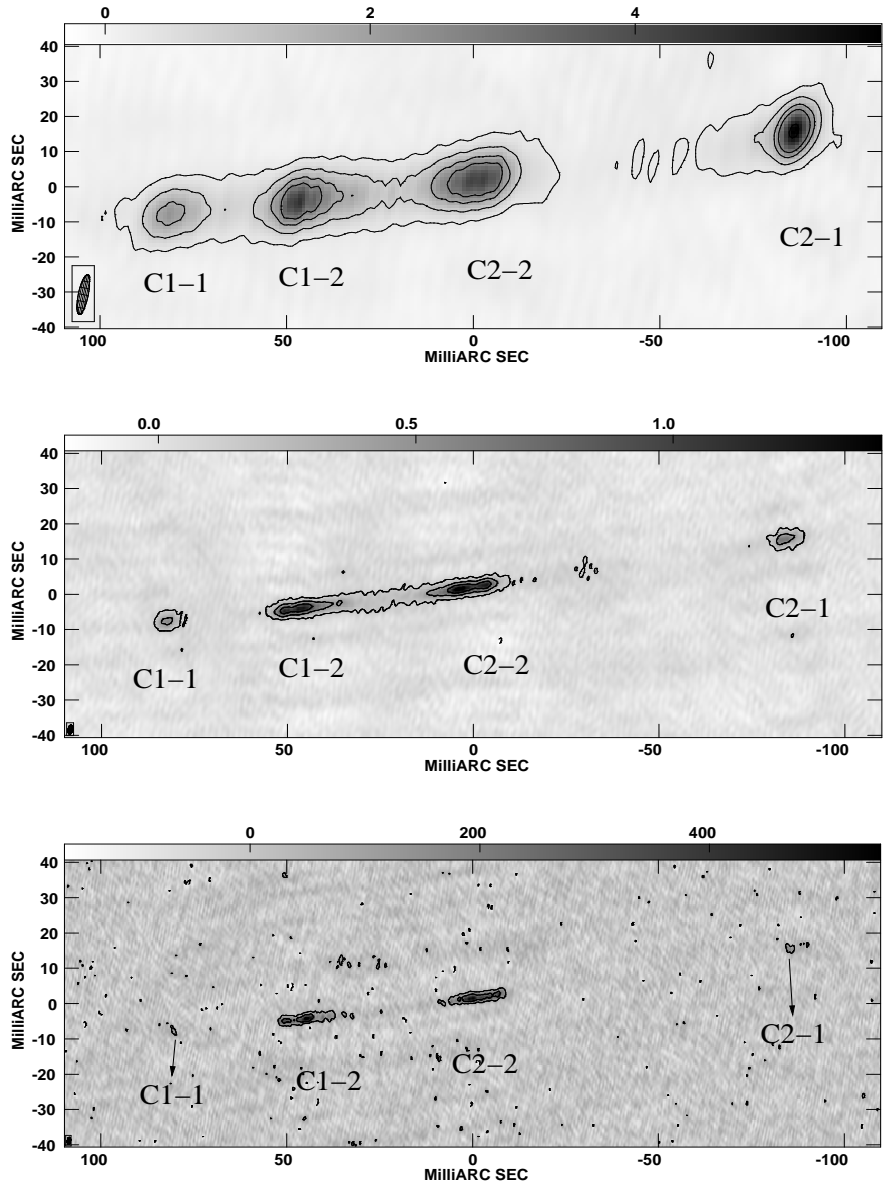}
   \end{center}
 \end{minipage}
\caption{Figures above show region C (partial images C1 and C2) at 18~cm
($\sigma$~=~0.1~mJy~beam$^{-1}$), 6~cm
($\sigma$~=~0.05~mJy~beam$^{-1}$) and 3.6~cm
($\sigma$~=~0.03~mJy~beam$^{-1}$) from top to bottom. The contours are
(-3, 3, 9, 18, 27, 54) $\times$ $\sigma$ in the respective maps.
The restoring beams are as for Figure~2.
}
\end{figure}
\begin{figure}
\begin{minipage}[]{0.5\textwidth}
  \begin{center}
   \includegraphics[height=9.5cm,width=7.3cm,angle=-90]{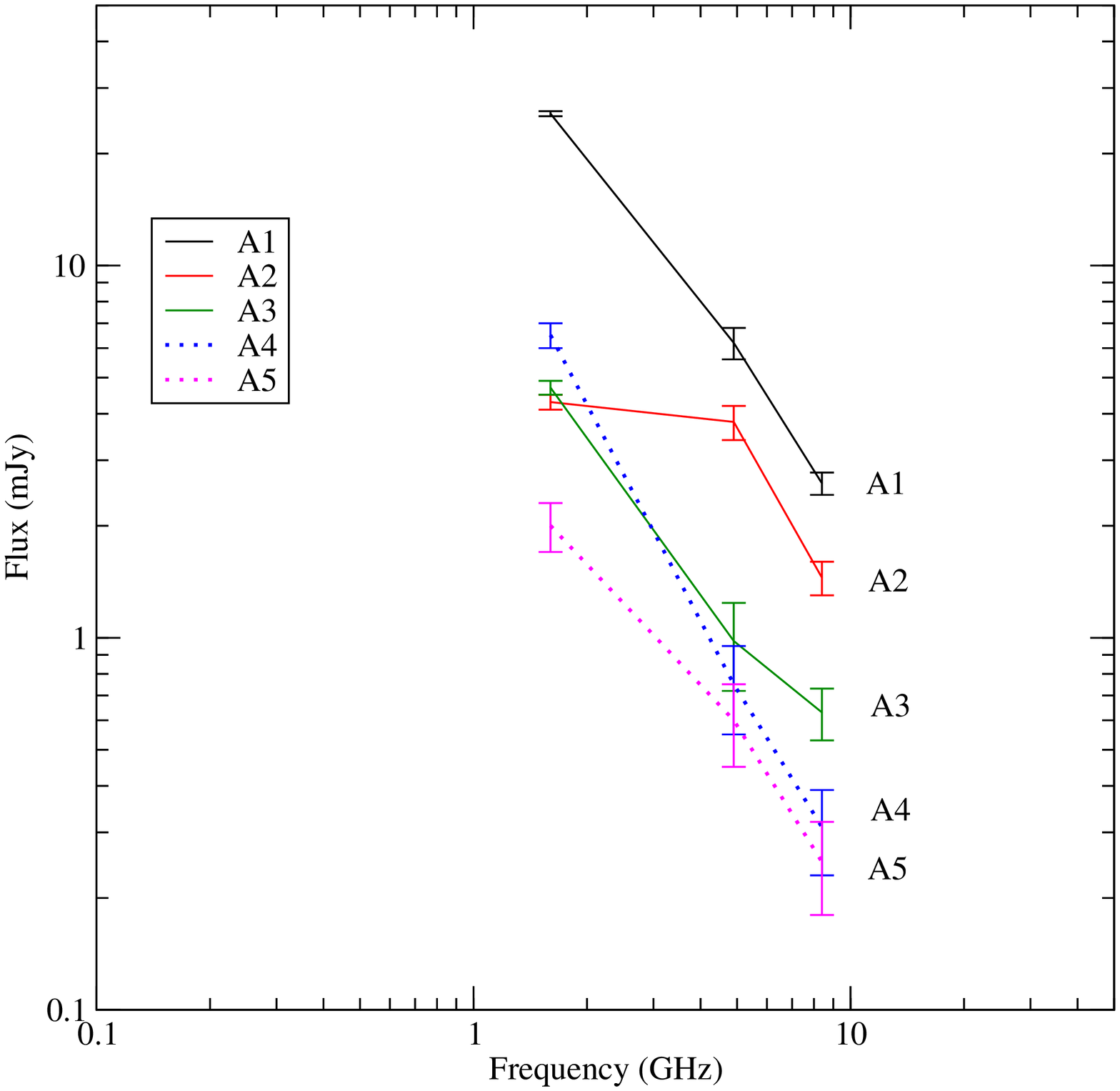}
  \end{center}
 \end{minipage}
\hfill
 \begin{minipage}[]{0.5\textwidth}
  \begin{center}
   \includegraphics[height=9.5cm,width=7.3cm,angle=-90]{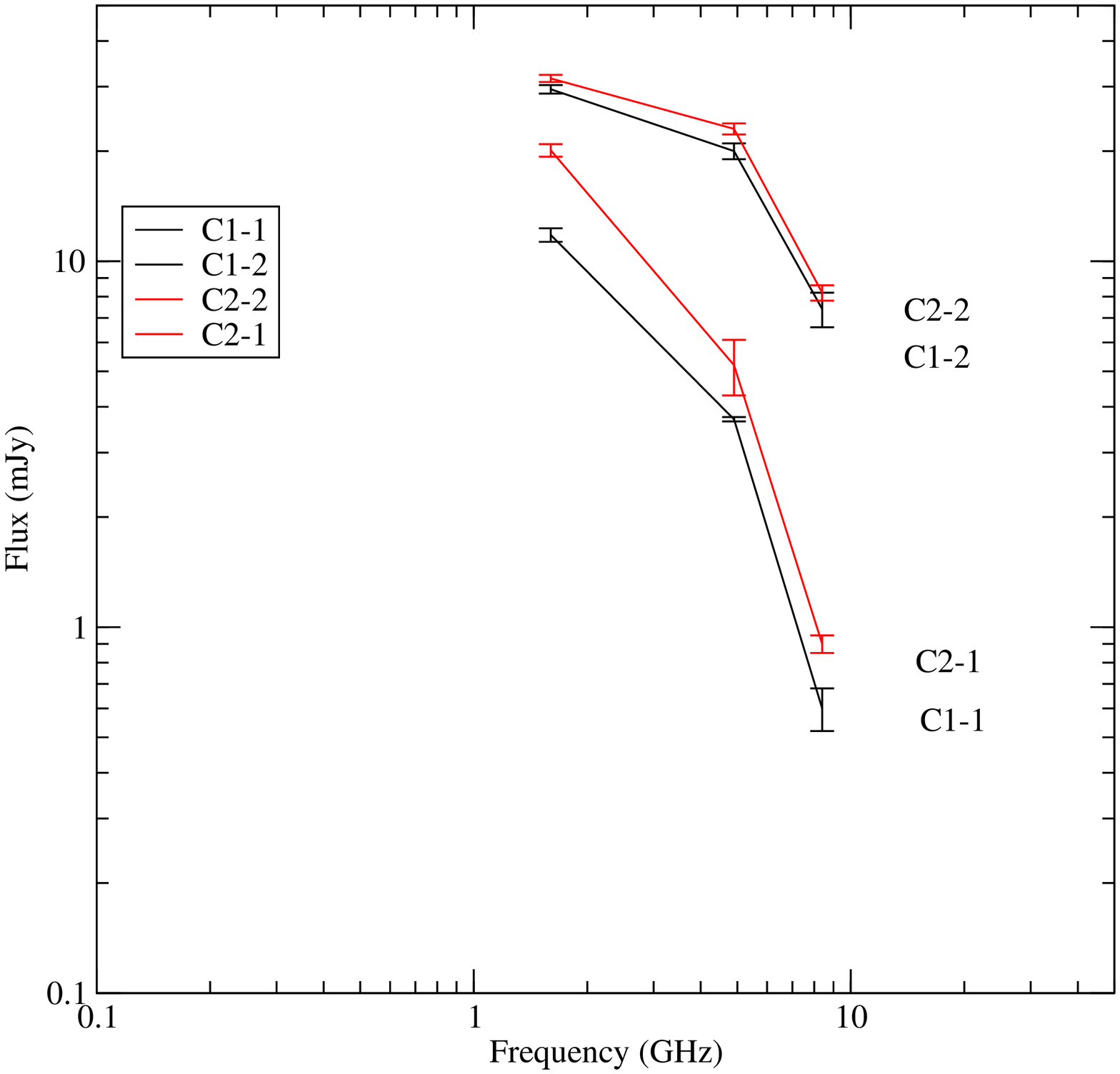}
  \end{center}
 \end{minipage}
\caption{Spectra of all components. \textit{Left}: Image A;
     \textit{Right}: Region C (images C1, C2) }
\end{figure}
and B2 as the ``core'' of the radio structure, and expect this to coincide
with the optical nucleus.

In the scenario of Koopmans et al, C1 and C2 are images of components
in the background source which appear in the NW end of image A. The
similar spectra of C2-2 (C1-2) and A2 (B2) suggests an identification
of C2-2 with A2, and hence A5 with C2-1. As noted by Koopmans,
however, models predict that C2-2 and C1-2 would be much more
magnified than they appear, and the coincident optical nucleus would
appear in C1 and C2. However, the data is consistent with a scheme
whereby the diamond caustic passes $\it{through}$ A2, causing only a
small part (not containing the small optical nucleus) to be quadruply
imaged.
From the gap between images C1 and C2 we deduce that A2 in
fact consists of a line of smaller, flat-spectrum subcomponents, the
caustic passing between 2 of them.  Alternatively, C1 and C2 may
correspond to further, faint components as yet undetected to the NW of
A; however, this would require there to be 2 flat-spectrum, core-like
components in the background source, which seems unlikely.
\acknowledgments
This work was supported by the European Community's Sixth Framework
Marie Curie Research Training Network Programme, Contract No.
MRTN-CT-2004-505183 "ANGLES".

\end{document}